\documentclass[aps,prd,twocolumn,superscriptaddress,amsfont,graphicx,nofootinbib,preprintnumbers]{revtex4}%
\usepackage{color,graphicx,epsfig}
\usepackage{ifpdf}
\usepackage{amsmath}
\usepackage{bm}
\usepackage{color}
\usepackage[english]{babel}
\usepackage{graphicx}%
\usepackage{amsfonts}%
\usepackage{amssymb}
\usepackage{braket}
\usepackage{hyperref}

\definecolor{nicered}{rgb}{0.7,0.1,0.1}
\definecolor{nicegreen}{rgb}{0.1,0.5,0.1}
\hypersetup{colorlinks,citecolor= nicegreen,linkcolor= nicered}

\newcommand{\beq}{\begin{equation}}
\newcommand{\eeq}{\end{equation}}
\newcommand{\bea}{\begin{eqnarray}}
\newcommand{\eea}{\end{eqnarray}}

\begin{document}

\def\LjubljanaFMF{Faculty of Mathematics and Physics, University of Ljubljana,
 Jadranska 19, 1000 Ljubljana, Slovenia }
\def\LjubljanaIJS{Jo\v zef Stefan Institute, Jamova 39, 1000 Ljubljana, Slovenia}
\def\CERN{CERN, Theory Division, CH-1211 Geneva 23, Switzerland}
\def\Firenze{INFN, Sezione di Firenze, Via G. Sansone, 1; I-50019 Sesto Fiorentino, Italy}

\title{Back to 1974: The $\mathcal Q-$onium }

\author{Jernej F.\ Kamenik} 
\email[Electronic address:]{jernej.kamenik@cern.ch} 
\affiliation{\LjubljanaIJS}
\affiliation{\LjubljanaFMF}

\author{Michele Redi} 
\email[Electronic address:]{michele.redi@fi.infn.it} 
\affiliation{\Firenze}

\preprint{...}

\date{\today}
\begin{abstract}
We show that the 750 GeV di-photon excess could be explained by the ${\mathcal Q}-$onium system 
of a new QCD-like theory with  fermions vectorial under the SM. Beside the spin-0 di-photon singlet this scenario predicts almost degenerate colored
scalars and spin-1 resonances analogous to the $J/\Psi$ in QCD. All these states are within the reach of the LHC. An apparent large width can be explained 
as due to production of excited states with splitting $\Delta m\sim \Gamma$.
\end{abstract}

\maketitle

\begin{figure}[t]
\centering
\includegraphics[width=1.0\hsize]{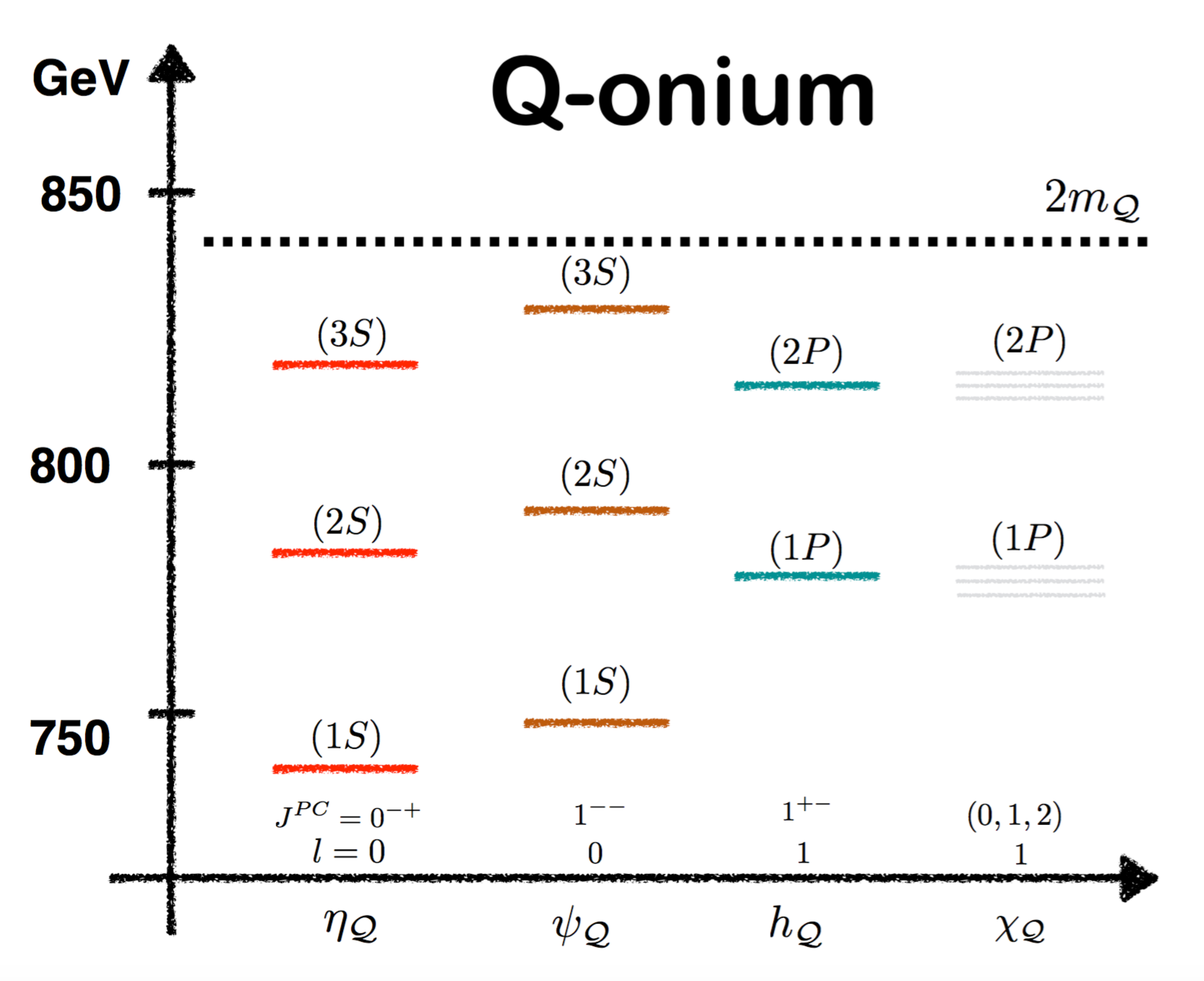}
\caption{\it Sketch of the $\mathcal Q$-onium system. The spectrum resembles the one of charmonium and bottomonium in QCD.  The di-photon resonance is interpreted
as the lightest singlet  ${}^1S_0$. Resonances of different spin are predicted and higher level excitations could account for an apparent large 
width of the resonance.}\label{fig:qonium}
\end{figure}

\section{Introduction}

A very plausible explanation of the di-photon excess at $M\simeq750$ GeV recently reported at the LHC~\cite{exp} is provided by new confining gauge dynamics, dubbed technicolor (TC), with fermions $\mathcal Q$ that are
vectorial under the SM~\cite{Franceschini:2015kwy,Craig:2015lra,others,Redi:2016kip}. Most theoretical speculations have focused on the regime 
where  the $\mathcal Q$ are lighter than the confinement scale. The di-photon resonance is then identified with a TC pion SM singlet that couples to SM gauge bosons through anomalies. 

In this letter we study the regime where the $\mathcal Q$ are heavier than the confinemant scale $\Lambda_{\rm TC}$, see \cite{Franceschini:2015kwy,Craig:2015lra,Agrawal:2015dbf,Curtin:2015jcv} for early work. The system so obtained is entirely analogous to quarkonium  in QCD, bound states of $c\bar{c}$ or $b\bar{b}$, see \cite{Novikov:1977dq,Appelquist:1978aq, Brambilla:2004} for a review. The di-photon resonance is identified with $\eta_{\mathcal Q}^1$,
the lightest spin-0 color singlet (${}^1S_0$) bound state. The analogous resonance in QCD, $\eta_c$ has a rate $\Gamma{(\eta_c \to \gamma\gamma)}/m_{\eta_c} \sim 2 \times 10^{-6}$ which is almost exactly what is required to reproduce  the di-photon excess~\cite{Franceschini:2015kwy}. A model independent prediction is the existence of a scalar color octet ($\eta_{\mathcal Q}^8$),
almost degenerate in mass and  coupled to pairs of gluons~\cite{Craig:2015lra}  as well as spin-1 excitations similar to the $J/\Psi$.

It is not too surprising that $\eta_{\mathcal Q}^1$ is the first resonance discovered at LHC. 
Indeed it is the lightest resonance that couples to gluons and photons. The almost degenerate $\eta_{\mathcal Q}^8$ is also copiously produced 
but it can only decay to jets and should be discovered in the next run of the LHC. If the interpretation given in this letter is correct, various resonances of the $\mathcal Q-$onium will be within the reach of the LHC. 

\section{Spectroscopy of $\eta_{\mathcal Q}$ and $\psi_{\mathcal Q}$}

For concreteness we consider $SU(N_{\rm TC})$ gauge theories with fermions $\mathcal Q$ in the fundamental representation.
Under the SM they form a vectorial representation $R$ (in general reducible) of the SM. The strong dynamics confines at the scale $\Lambda_{\rm TC}<m_{\mathcal Q}$ producing $\mathcal Q \bar{\mathcal Q}$ bounds states analogous to the charmonium with mass $M\sim 2 m_{\mathcal Q}$. The expected spectrum is sketched in Fig.~\ref{fig:qonium}.
For each strong dynamics level we obtain a fine structure of SM multiplets. In order for the resonances to couple to gluons and photons the
new fermions should carry color and electric charge.  A SM rep $R=(d_3,d_2)_Y$ will then produce ${\mathcal Q}$-onium states as,
\begin{equation}
R\times \bar{R}= (1,1)_0 + (8,1)_0 + \dots \,,
\end{equation}
where the elipses denote possible further representations, depending on $R$. The presence of the di-photon singlet coupling to photons and gluons is always accompanied by a scalar color octet.
The ground state spin-zero (${}^1S_0$) color singlet ($\eta_{\mathcal Q}^1$) decays to gluons and photons and will be identified 
with the di-photon resonance while the spin-0 color octet ($\eta_{\mathcal Q}^8$) couples only to gluons. 
Spin-1 states, (${}^3S_1$) color singlet ($\psi_{\mathcal Q}^1$) and color octet  ($\psi_{\mathcal Q}^8$),  couple instead to pairs of SM fermions or 3 SM gauge bosons.

Other colored states could appear for example $(8,3)_0$ that would  couple to $W$ bosons and gluons. 
If other fermions with mass above $\Lambda_{\rm TC}$ exist more $\mathcal Q$-onium bound  states will be formed. 
One difference with the TC pion  scenario ($\Lambda_{TC} > m_{\mathcal Q}$) is that these states will not significantly mix unless the masses are almost 
degenerate so they will appear as separate resonances. Therefore we can focus on irreducible SM reps in what follows.

The dynamical scale of the theory is given by,
\begin{equation}
\Lambda_{\rm TC}\sim M \exp\left[-\frac {6\pi}{(11 N_{\rm TC}-2 n) \alpha_{\rm TC}(M)}\right]\,,
\end{equation}
where we have included $n$ light flavors. For $\alpha_{\rm TC}> \alpha_s$  the bound states are TC singlets formed due to the new strong interactions \cite{alphascale}.
Two regimes can be distinguished. If $\alpha_{\rm TC} M \gg \Lambda_{\rm TC}$ confinement gives small 
corrections and the system can be described in first approximation as a positronium-like bound state with Coulomb potential $V=- C_N{\alpha}_{\rm TC}/r$. 
For fermions in the fundamental rep  $C_N=(N^2-1)/(2N)$. The binding energies are given by,
\begin{equation}
\Delta E^{(n,l)}_{\rm Coul.}= -\frac {C_N^2 {\alpha}_{\rm TC}^2} {8 n^2}  M\,,~~~ 
\label{eq:hydrogen}
\end{equation}
where $n=1,2,\ldots$ are the radial excitation levels and $\alpha_{\rm TC}$ should be evaluated at the scale of the bound state size $\sqrt{\langle r^2\rangle} \ll 1/\Lambda_{\rm TC}$. For this we need the information on the radial wave-function  $R(r)$, normalized to $\int_0^\infty |R(r)|^2 r^2 dr = 1$.  In what follows we will be primarily interested in its value  at the origin ($|R^{}(0)|$) . In the Coulomb regime for the $n-$th radial excitation this is given by,
\begin{equation}
\left(\frac{|R^{(n,l)}(0)|^2}{M^3}\right)_{\rm Coul.}  = \frac 1 {16 n^3} (C_N {\alpha}_{\rm TC})^3 \,.
\end{equation}
We note that  such a weakly coupled picture fails if applied to the lowest lying states of charmonia and bottomonia, and a recent numerical lattice QCD simulation indicates deviations from  the positronium-like behavior even for QCD bound states with mass close to $M$~\cite{Kim:2015zqa}, even though $\alpha_s(M) M \gg  \Lambda_{\rm QCD}$.

In the opposite regime $\alpha_{\rm TC} M \ll \Lambda_{\rm TC}$ the confinement effects modify significantly the bound state and splitting of energies becomes larger. 
Moreover while $|R(0)|^2/M^3$ is constant in the Coulomb regime, $|R(0)|^2$ becomes almost independent of $M$ when confinement effects dominate. 
In Table \ref{table:QCDeta} we report the masses and wave-function values extracted for 
QCD $\eta$ singlets. For both charmonium and bottomonium, confinement effects appear to be dominant.

\begin{table}[t]
\begin{tabular}{c||c|c|c|c}
 $\eta_X$ & $\displaystyle{m_{\eta_X}[\rm GeV]}$ & $\displaystyle{\frac{\Gamma(\eta_X\to {\gamma\gamma})} {m_{\eta_X}}}$  &~$\displaystyle{\frac{|R(0)|^2}{m_{\eta_X}^3}}$  & \\  \hline
$\eta'$ & 0.958   & $5\times 10^{-6}$ & --  & \\
$\eta_c(1S)$ & 2.983 & $2\times 10^{-6}$ & $1.5 \times 10^{-2}$&  \\
$\eta_c(2S)$ & 3.639 & $10^{-6}$ & $6 \times 10^{-3}$ &  \\
$\eta_b(1S)$ & 9.398 & $5\times 10^{-8}$ & $6 \times 10^{-3}$ & \\
$\eta_b(2S)$ & 10 & $2\times 10^{-8}$ & $2.5 \times 10^{-3}$ & \\
\end{tabular}
\caption{\em $\eta_{\mathcal Q}^1$ singlets in QCD \cite{PDG}. Their widths into photons, are not  measured directly, but are derived using the decay of 
$\psi$ into electrons through $\Gamma(\psi_{\mathcal Q} \to \bar f f)/\Gamma(\eta_{\mathcal Q} \to \gamma\gamma)=Q_f^2/(3 Q_{\mathcal Q}^2)$.  
The value of the wave-function at the origin is extracted using the formula 
$\Gamma(\eta_{\mathcal Q} \to \gamma\gamma) = 12  \alpha^2 Q_{\mathcal Q}^4 |R(0)|^2/M^2$. }
\label{table:QCDeta}
\end{table}

In the  perturbative $ \alpha_{TC}$ regime the mass splitting between the ${}^1 S_0$ and ${}^3 S_1$ states can be estimated 
analogously to the hyperfine structure of positronium or atoms
{\beq
\left(\frac{\Delta M} M\right)_{\rm HF} =  \frac{8}{3} C_N {\alpha}_{\rm TC} \frac{|R(0)|^2}{M^3} \stackrel{\rm Coul.}{ =} \frac{1}{6n^3}(C_N \alpha_{\rm TC})^4\,.
\eeq}
where the second equality is valid in the Coulomb-like limit (when $\alpha_{\rm TC} M \gg \Lambda_{\rm TC}$). The mass splitting is thus extremely sensitive to the precise value of $\alpha_{\rm TC}$. This could be expected, since in the chiral regime ($m_{\mathcal Q} \ll \Lambda_{\rm TC}$) ${}^1S_0$ state becomes a Nambu-Goldstone boson of the approximate chiral $\mathcal Q$ flavor symmetry, while in the asymptoticly free $m_{\mathcal Q} \to \infty$ limit  (when TC interactions are not strong enough to flip the spin of $\mathcal Q$), spin becomes a globally conserved quantum number of the TC sector.
For the charmonium this splitting if  3.7\%, and for bottomonium 0.7\%\,.

Within each $\mathcal Q$-onium level SM interactions split the multiplets. Assuming that the
bound state is formed due to the TC interactions this can be treated as a small perturbation and implies that the splitting  
is linear in $\alpha_s$
\begin{equation}
(\Delta M)_{QCD}   \sim C_3  \alpha_s\, |R(0)|^{\frac 2 3}\,.
\label{colorsplitting}
\end{equation}
When $\alpha_{\rm TC}$ becomes comparable with $\alpha_s \approx 0.1$ it is important to include QCD effects for the bound state.
For the singlet this provides an extra attractive force so that in the Coulomb regime the effective coupling that controls the bound state is replaced by $C_N \alpha_{\rm TC}+ C_3 \alpha_s$. {We note that recent calculations using potential models and lattice simulations estimate the irreducible QCD contribution to $(|R_1(0)|^2/M^3)_{\rm QCD} \sim (0.0002 - 0.0008)$~\cite{Hagiwara:1990sq, Kim:2015zqa}.} For the color octet combination QCD is repulsive so that this state is more loosely bound, the effective becoming $C_N \alpha_{\rm TC} + (C_3 - 3/2) \alpha_s$. For example for $\mathcal Q$ in the fundamental of both TC and QCD and $N_{\rm TC}=3$ the effective coupling of the octet at  $\alpha_{\rm TC} \approx \alpha_s$ is reduced by $\approx 60\%$ compared to the singlet, leading to $|R_8(0)|/|R_1(0)| \approx 0.3$.

Finally, electro-weak interactions split the components of $\mathcal Q$ $SU(2)$ multiplets. For $\mathcal Q$ doublets for example
 $\Delta m_{\mathcal Q}= \alpha_2 Y m_W s_W^2/c_W\approx 0.7\,{\rm  GeV}\times Y$ \cite{Cirelli:2005uq}.  When this is smaller than the width of the bound states ($\Gamma$), $\mathcal Q$-onium states will fill complete $SU(2)$ multiplets with little mixing between them. Only the $SU(2)$ singlet states can couple to gluons in this case and the rates to electro-weak final states are identical to the analogous TC pion scenario, see~\cite{Redi:2016kip}.
In the opposite regime $\Delta m_{\mathcal Q}/\Gamma > 1$,  the $\mathcal Q$-onium mass eigenstates will be aligned with the fermion $\mathcal Q$ charge eigenstates, which must thus be summed incoherently. In this regime, the relative decay rates into electro-weak final states do not follow from $SU(2)$ relation of the EFT~\cite{Kamenik:2016tuv}. In particular, the di-photon excess will be dominated by the bound state made by the $SU(2)$ component of $\mathcal Q$ with the highest electric charge. As we will see, the first regime is relevant for the $\mathcal{Q}-$onium system made of $Q=(3,2)_{1/6}$ while the latter holds for $Y=(3,2)_{-5/6}$ when $\eta^1_{\mathcal Q}$ width is dominated by decays to gluons.

\section{$\mathcal Q-$onium production and decays at LHC}

Since in the heavy $\mathcal Q$ limit ${}^1S_0$ and ${}^3 S_1$ states are related by spin-symmetry, on can relate all their dominant interactions purely in terms of SM gauge group invariants and charges.  
We choose  $\eta_{\mathcal Q}^1 \to gg$ decay as the reference width. One finds,
\beq
\frac{\Gamma(\eta_{\mathcal Q}^1 \to gg)}M = 32  N_{\rm TC} d_2 \frac {I_3^2}{d_3} \alpha_s^2 \frac{|R_1(0)|^2}{M^3}\,,
\label{eq:widthgluons}
\eeq
where $I_3=1/2$ for the fundamental representation. For $\Delta m_{\mathcal Q}/\Gamma>1$ the formula above applies for each $SU(2)$ component with $d_2=1$.

The $\eta_{\mathcal Q}$  decay to gluon or (only for $\eta^1_{\mathcal Q}$) photon pairs, while $\psi_{\mathcal Q}$ decay to pairs of fermions or three gauge bosons.
The prompt single production of $\eta^{1,8}_{\mathcal Q}$ states at the LHC then proceeds dominantly through gluon fusion, 
while $\psi_{\mathcal Q}^{1,8}$ are produced via $q\bar q$ annihilation. 
{For $\Delta m_{\mathcal Q}/\Gamma >1$} {(relevant for the models $U$, $X$ and $Y$ in table \ref{table:models})} the decay rates into SM states for the various components of a $\mathcal Q$ $SU(2)$ multiplet are predicted as,
\begin{align}
\frac {\Gamma(\eta_{\mathcal Q}^1 \to \gamma\gamma)}{\Gamma(\eta_{\mathcal Q}^1 \to gg)} & = \frac {d_3^2}  8 \frac{Q_{\mathcal Q}^4}{I_3^2}\frac{\alpha^2}{ \alpha_s^2} \,,\nonumber  \\
\frac {\Gamma(\eta_{\mathcal Q}^1 \to Z\gamma)}{\Gamma(\eta_{\mathcal Q}^1 \to gg)} & = \frac {d_3^2}  {4 s_W^2 c_W^2} \frac{Q_{\mathcal Q}^2 (T^3_{\mathcal Q}-s_W^2 Q_{\mathcal Q})^2}{I_3^2}\frac{\alpha^2}{ \alpha_s^2} \,,\nonumber  \\
\frac {\Gamma(\eta_{\mathcal Q}^1 \to ZZ)}{\Gamma(\eta_{\mathcal Q}^1 \to gg)} & = \frac {d_3^2}  {8 s_W^4 c_W^4} \frac{(T^3_{\mathcal Q}-s_W^2 Q_{\mathcal Q})^4}{I_3^2}\frac{\alpha^2}{ \alpha_s^2} \,,\nonumber  \\
\frac{\Gamma(\psi_{\mathcal Q}^1 \to f\bar f)}{\Gamma(\eta_{\mathcal Q}^1 \to gg)} & =\frac {d_3^2} {48\,s_W^4\,c_W^4} \left(\frac{c_W^2\,T^3_{\mathcal Q}T^3_f +s_W^2\, Y_{\mathcal Q} Y_f}{I_3}\right)^2 \frac{\alpha^2}{ \alpha_s^2} \,, \nonumber \\
\frac{\Gamma(\eta_{\mathcal Q}^8 \to gg)}{\Gamma(\eta_{\mathcal Q}^1 \to gg)} & = \frac {d_3 D_3}{1024\,I_3^3} \frac{|R_8(0)|^2}{|R_1(0)|^2} \,, \nonumber \\
\frac{\Gamma(\eta_{\mathcal Q}^8 \to g\gamma)}{\Gamma(\eta_{\mathcal Q}^1 \to gg)} & = \frac {3 d_3 D_3}{640\,I_3^3} \frac{|R_8(0)|^2}{|R_1(0)|^2} \frac {\alpha\,Q_{\mathcal Q}^2}{\alpha_s} \,, \nonumber \\
\frac{\Gamma(\psi_{\mathcal Q}^8 \to q\bar q)}{\Gamma(\eta_{\mathcal Q}^1 \to gg)} & ={ \frac {d_3}{48 I_3}} \frac{|R_8(0)|^2}{|R_1(0)|^2} \,,
\label{eq:ratios}
\end{align}
where $f$ refers to a single flavor of chiral fermions, $T^3_{\mathcal Q}$ is the third component of the weak isospin, ${\rm Tr}[T^a T^b]=I \delta^{ab}$  and $D=\sum_{abc} d_{abc}^2 $ with $d_{abc}=2{\rm Tr}[T_a\{T_b, T_c\}]$. Furthermore for color triplets $D_3=40/3$.  The values of the wave-function at the origin for singlet and octet combinations differ due to QCD effects so that $|R_8(0)|<|R_1(0)|$ while we neglect the splitting between spin-0 and spin-1 states due to TC interactions. 
The $\psi$ decay widths into 3 SM gauge bosons are also predicted. For example the width of $\psi_{\mathcal Q}^1$ into 3 gluons reads,
\begin{align}
\frac{\Gamma(\psi_{\mathcal Q}^1 \to g g g)}{\Gamma(\eta_{\mathcal Q}^1 \to gg)} & = \frac {D_3} {288\, I_3^2}\frac{\pi^2-9}{\pi} {\alpha_s}\,.
\end{align}
which is extremely small. Many other relations can be found generalising the ones in \cite{Novikov:1977dq}.

The formulas above can be easily adapted to the degenerate $SU(2)$ limit ($\Delta m_{\mathcal Q}/\Gamma <1$).
The ones with electro-weak gauge bosons final states can be read from~\cite{Redi:2016kip}.
For $\eta_{\mathcal Q}^8 \to g \gamma$ only the hypercharge contributes in eq. (\ref{eq:ratios}). {For the $\psi_{\mathcal Q} \to f\bar f$ one has 
\begin{align}
\frac{\Gamma(\psi_{\mathcal Q}^{(1,1)} \to f \bar f)}{\Gamma(\eta_{\mathcal Q}^1 \to gg)} & =\frac {d_3^2} {48\,c_W^4} \left(\frac{Y_{\mathcal Q}\,Y_f }{ I_3}\right)^2 \frac{\alpha^2}{ \alpha_s^2} \,, \nonumber \\
\frac{\Gamma(\psi_{\mathcal Q}^{(1,3)} \to f_L\bar f_L)}{\Gamma(\eta_{\mathcal Q}^1 \to gg)} & =\frac {d_3^2} {48\,s_W^4} \frac{I_2}{2 d_2 I_3^2} \frac{\alpha^2}{ \alpha_s^2} \,, 
\end{align}
where $\psi_{\mathcal Q}^{(1,1)}$ and $\psi_{\mathcal Q}^{(1,3)}$ refer to the $SU(2)$ singlet and the neutral component of the triplet,  (both color singlets) respectively. }The other ratios are not modified. 
Considering $\eta_{\mathcal Q}^1 \to ZZ, Z\gamma$ decays and assuming a signal cross-section $\sigma(pp\to \gamma\gamma)\approx 5$ fb~\cite{Kamenik:2016tuv} at LHC 13, the bounds from run 1~\cite{Aad:2015kna, Khachatryan:2015cwa} and recently run 2~\cite{Zgamma} translate into a constraint on the dimension of the $SU(2)$ representation ($d_2$) and hypercharge ($Y$)
\begin{equation}
-3.5 < \frac {d_2^2-1}{Y^2}< 30\,.
\end{equation}
Finally, we note in passing that for $\mathcal Q$ in non-trivial $SU(2)$ representations, charged $\eta$ and $\psi$ states will also be formed.
However, they are singly produced  only through weak interactions and thus less relevant for LHC phenomenology. 

In the narrow width approximation, the resonant production cross-sections of $\eta_{\mathcal Q}$ and $\psi_{\mathcal Q}$ are given by,
\begin{equation}
 \sigma(pp\to X) =\frac{(2 J_X+1) D_X}{M s} 
\sum_{\mathcal P} C_{{\mathcal{P P}}} K^X_{{\mathcal{P P}}} \Gamma(X\to \mathcal {P P}) \,,
\end{equation}
where $D_X$ is the dimension of the representaton, $J_X$ the spin and $\mathcal P$ is the parton producing the resonance at the LHC: gluons for $X=\eta_{\mathcal Q}$ and quarks for $X=\psi_{\mathcal Q}$.
The  parton luminosity coefficients at LHC 13(8) for the production of a 750 GeV resonance in the s-channel are $C_{gg}=2137(174)$, $C_{u\bar{u}}=1054(158)$ and $C_{d\bar{d}}=627(89)$~\cite{Franceschini:2015kwy}.
In our phenomenological analysis we also include (approximately) known NLO QCD $K-$factors of $K^{\eta^1}_{gg}=1.6$~\cite{Harlander:2005rq},  $K^{\eta^1}_{q\bar q}=1.2$~\cite{Drees:1989du}, $K^{\psi^1}_{q\bar q}=1.3$~\cite{Fuks:2007gk} and $K^{\psi^8}_{q\bar q}=1.3$~\cite{Chivukula:2011ng}. On the other hand, QCD corrections to prompt production of a massive color octet scalar are presently not known. In Table \ref{table:models} we take $K^{\eta^8}_{gg}= K^{\eta^1}_{gg}$, consistent with results in~\cite{Idilbi:2009cc} considering a somewhat similar scenario, but our results can easily be rescaled for different values. 

\begin{table}[t]
\begin{tabular}{c||c|c|c|c|c}
& $\sigma_{U}$[fb] & $\sigma_{X}$[fb]  & $\sigma_{Q}$[fb]  & $\sigma_{Y}$[fb]  & $ {\sigma{\rm [fb]}}{}$\\
   \hline
$pp \to \eta^1 \to gg$ & 200 & 12 & 500 & 25  & $<$2500\\
$pp \to \eta^1 \to \gamma Z$ & 0.6  & 0.6 & {5} & 0.4  &$<11$\\
$pp \to \eta^1 \to ZZ$ & 0.1 & 0.1 & {9} &1.2  &$<12$\\
$pp \to \psi^1 \to e \bar{e}$ & 0.3 & 0.07 & 1 & 0.1   & $<$1.2 \\
\hline
$pp \to \eta^8 \to gg$ & 500 & 30  & 1250 & 60  & $<$2500\\
$pp \to \eta^8 \to g \gamma$ & 80 & 20  & 13 & 20  & $<30$  \\
$pp \to \psi^8 \to jj$ & 600 & 35 & 1450 & 70  & $<$2500  \\
$pp \to \psi^8 \to t\bar t$ & 110 & 7 & 290  & 15  & $<$600  \\
\end{tabular}
\caption{\em Cross-sections for the $\mathcal Q$-onium system made of the $U=(3,1)_{2/3}$, $X=(3,1)_{4/3}$, $Q=(3,2)_{1/6}$ and $Y=(3,2)_{-5/6}$ fermions at the 8TeV LHC. 
We assume $\sigma(pp\to \eta^1 \to\gamma\gamma)=5$ fb at the 13TeV LHC and no invisible decays. The experimental constraints on di-jet, $\gamma$+jet, di-lepton, $ZZ$, $Z\gamma$ and $t\bar t$ resonances are taken from~\cite{CMS:2015neg, Aad:2014cka, Aad:2015fna, gammaj, Aad:2014fha, Aad:2015kna}. {For the $Y$ model, all values correspond to summed contributions of $\mathcal Q$ charge eigenstates. Similarly, for the $Q$ model, the $\psi^1$ label refers to the sum of $\psi^{(1,1)}$ and $\psi^{(1,3)}$ contributions.} The rates at LHC13 can be obtained multiplying by $r^{13/8}_{gg}\approx 4.7$ and $r^{13/8}_{q\bar{q}}\approx 2.5$  the rates at LHC8 of $\eta_{1,8}$ and $\psi_{1,8}$  respectively.}
\label{table:models}
\end{table}

The di-photon signal cross-section is reproduced for,
\begin{equation}
\frac {\Gamma(\eta_{\mathcal Q}^1 \to \gamma\gamma)}{M}\frac {\Gamma(\eta_{\mathcal Q}^1 \to gg)}{\Gamma} \approx 0.7\times  10^{-6}\,,
\label{eq:signal}
\end{equation}
implying that $\Gamma{(\eta^1_{\mathcal Q} \to \gamma\gamma)}/M\ge 0.7 \times 10^{-6}$ with the equality saturated when the  width is dominated by decays into gluons.

Let us consider models with an $SU(2)$ singlet $\mathcal Q$ in detail. The width into photons reads
\beq
\frac{\Gamma(\eta_{\mathcal Q}^1 \to \gamma\gamma)}M = 12 N_{\rm TC}  \alpha^2 Q^4 \frac{|R_1(0)|^2}{M^3}\,.
\label{eq:widthphotons}
\eeq
Reproducing the di-photon signal assuming that the total with is dominated by decays to gluons then requires,
\begin{equation}
\frac{|R_1(0)|^2}{M^3} \approx  10^{-3} \frac 1 {N_{\rm TC}\,Q^4}\,.
\label{eq:rate}
\end{equation}
or larger if extra decay channels exist.
Given the irreducible QCD contribution to $R_1(0)$~\cite{Hagiwara:1990sq}, eq. (\ref{eq:rate}) can only be satisfied for $Q\lesssim 0.5 (3/N_{\rm TC})^{1/4}$ in the  $\alpha_{\rm TC}\lesssim \alpha_s$ limit, suggesting the necessity of extra decay channels. These are naturally provided  by TC glueballs and lighter TC pions. 
For example in QCD ${\rm Br}(\eta_c\to \gamma \gamma)\sim 10^{-4}$ due to decays into hadrons.

Given that $\eta_{\mathcal Q}/\psi_{\mathcal Q}$ are almost degenerate, all other cross-sections are predicted in this model up to the difference between wave-function of singlets and octets. Using gluon and quark lumininosities at 13 TeV we find,
\begin{equation}
\frac {\sigma(pp\to \eta_{\mathcal Q}^8 \to gg )}{ \sigma(pp\to \eta_{\mathcal Q}^1 \to gg )}\approx 4 \frac {\sigma(pp\to \psi_{\mathcal Q}^8 \to j j)}{ \sigma(pp\to \eta_{\mathcal Q}^1 \to gg)} \approx 2.5 \frac{|R_8(0)|^2}{|R_1(0)|^2} \,,
\end{equation}
where jets from $\psi_{\mathcal Q}^8$ include $b$-quarks. Consequently, for $ \alpha_{\rm TC} \gg \alpha_s$,  $\eta_{\mathcal Q}^8$ gives the dominant contribution to the resonant di-jet cross-section at LHC 13. At LHC 8 instead $\eta_{\mathcal Q}^8$ and $\psi_{\mathcal Q}^8$ give comparable di-jet signals.
Estimates for various representations are given in Table \ref{table:models} assuming no extra decay channels and equal wave-functions for singlet and octets. 
Note that in this regime $\sigma(pp\to \eta_{\mathcal Q}^8 \to g\gamma)$ typically provides the strongest experimental constraint~\cite{gammaj}. 
For all color octet rates however these estimates should be taken as a conservative upper bound given that the QCD effects make the octets more weakly bound. 
A comparison with bounds at the 8 TeV LHC~\cite{CMS:2015neg,gammaj} including contributions from $\eta_{\mathcal Q}^1$, $\psi_{\mathcal Q}^1$ (and $\eta_{\mathcal Q}^8$, $\psi_{\mathcal Q}^8$ with $R_8(0) \simeq R_1(0)$) for the $SU(2)$ singlet $\mathcal Q$ implies,
\begin{equation}
Q> 0.3(1.0)\,,
\end{equation}
for the case $|R_8(0)| \ll |R_1(0)|$ ($|R_8(0)| \simeq |R_1(0)|$)
, respectively. 
Thus $\mathcal Q$ with the SM quantum numbers of the right-handed down quarks ($D$) and up-quarks ($U$) are disfavored only in the deeply bound regime where QCD effects are negligible. In the weakly bound regime where color octet effects are subleading, $D$ is only marginally compatible with existing di-jet bounds.

\begin{figure}[t]
\centering
\includegraphics[width=.95\hsize]{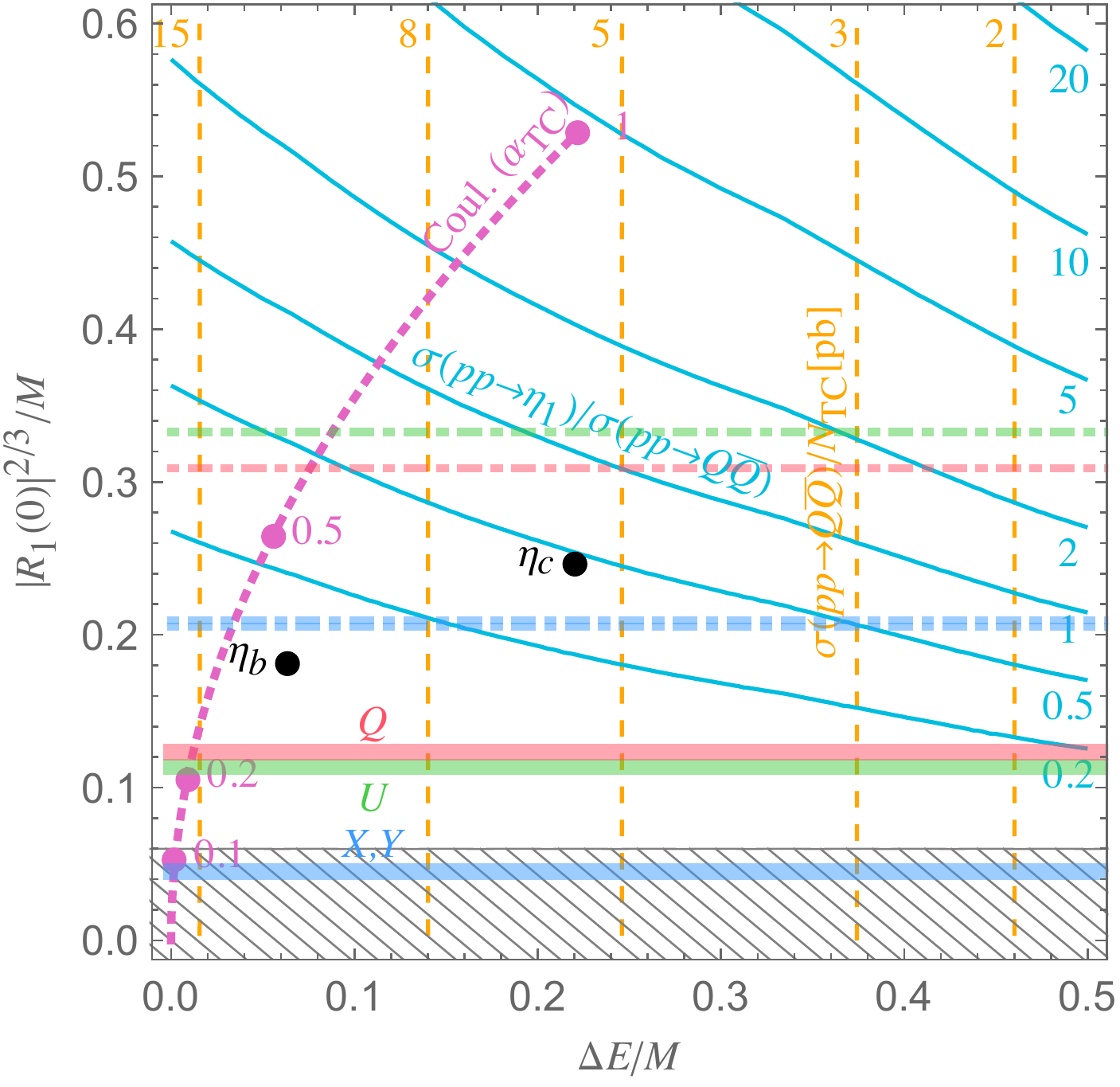}
\caption{\it Prompt $\eta_{\mathcal Q}^1$ vs. continuum $Q\bar{Q}$ production at 13 TeV centre of mass energy. Vertical lines (orange dashed) are contours
of constant $\sigma(pp \to \mathcal Q \bar{\mathcal Q})$ computed at NNLO in QCD~\cite{Czakon:2013goa} and normalized to $N_{\rm TC}$ while blue full contours correspond to the ratio $\sigma(pp \to \eta^1_{\mathcal Q})/\sigma(pp \to \mathcal Q \bar{\mathcal Q})$  as a function of $\Delta E/M$ and $|R_1(0)|^{2/3}/M$.  
Black dots correspond to the values of $|R_1(0)|$ and $\Delta E$ for  $\eta_c$ and $\eta_b$ in QCD. The horizontal 
contours correspond to the values of $|R_1(0)|$ reproducing the LHC di-photon excess for the cases of $U= (3,1)_{2/3}$ (shaded in green), $Q=(3,2)_{1/6}$ (shaded in red) and overlapping $X=(3,1)_{4/3}$, $Y=(3,2)_{-5/6}$ (shaded in blue), taking for concreteness $N_{\rm TC}=3$ and assuming predominantly prompt $\eta_{\mathcal Q}$ production and no additional significant decay modes (in full lines) or saturating their total decay width of $\sim 45$~GeV with hidden decay channels (in dot-dashed lines). 
For illustration we also report the values predicted in the Coulomb regime for various choices of $\alpha_{TC}$ for $N_{\rm TC}=3$\, (marked in purple points and connected with a dotted purple line), as well as an estimate~\cite{Hagiwara:1990sq} of the irreducible QCD contribution to $|R_1(0)|$ (upper edge of the black hashed region)\,.
}\label{fig:money}
\end{figure}

In addition to single prompt production, the lowest lying $\eta_{\mathcal Q}$ and $\psi_{\mathcal Q}$ states will also be produced in fragmentation of high $p_T$ QCD produced $\mathcal Q\bar {\mathcal Q}$ pairs (inclusive {\it continuum }production). In these processes, the confining TC dynamics forces formation of heavy highly excited $\bar {\mathcal Q}\mathcal Q$ bound states, which in turn decay to the lowest lying states ($\eta^{1,8}_{\mathcal Q}$ and $\psi^{1,8}_{\mathcal Q}$) by radiating TC glueballs, gluons and photons (see for example~\cite{Curtin:2015jcv}). 

Depending on the TC dynamics, either prompt or inclusive continuum production of $\eta_{\mathcal Q}$ may dominate. 
As an example we consider color triplet $\mathcal Q$. Both prompt  $\eta_{\mathcal Q}$ and continuum inclusive $\mathcal Q \bar{\mathcal Q}$ production are completely determined in terms of the $\eta_Q$ binding energy ($\Delta E$, or correspondingly $m_{\mathcal Q}$) and the $\eta_{\mathcal Q}$ radial wavefunction at the origin ($|R_1(0)|$). We compare the two in Fig.~\ref{fig:money}, where we plot $\sigma(pp \to \mathcal Q \bar{\mathcal Q})$ computed at NNLO in QCD~\cite{Czakon:2013goa} and normalized to $N_{\rm TC}$ (in orange dashed contours) as well as the ratio $\sigma(pp \to \eta^1_{\mathcal Q})/\sigma(pp \to \mathcal Q \bar{\mathcal Q})$ (in blue full contours) both at 13 TeV. We observe that in the tightly bound regime  ($\Delta E /M \sim \mathcal O(1)$ and/or $|R_1(0)|^2/M^3 \sim \mathcal O(1)$) prompt  $\eta_{\mathcal Q}$ production can easily dominate over continuum $ \mathcal Q \bar{\mathcal Q}$ production. However even in the weakly bound Coulomb regime, when continuum production is bigger, it may not necessarily be the dominant source of $\eta_{\mathcal Q}$'s, since only a (small) fraction of $\mathcal Q \bar{\mathcal Q}$ events will result in a specific $\eta_{\mathcal Q}$ state. 

On the same plot we also show the combinations of $\Delta E$ and $|R_1(0)|$ as predicted in the Coulomb limit (drawn with purple dotted line) and mark various values of $\alpha_{TC}$ for the case $N_{\rm TC}=3$\, (in purple points). This can be compared with the experimentally determined values for the case of $\eta_c$ and $\eta_b$ (marked in black points), where their binding energies are approximated by the mass differences between the $1S$ and $2S$ states in Table~\ref{table:QCDeta}\,. 
Finally, we overlay the values of $|R_1(0)|^{2/3}$ reproducing the LHC di-photon excess at $N_{\rm TC}=3$, for the cases of $U= (3,1)_{2/3}$ (shaded in green), $Q=(3,2)_{1/6}$ (shaded in red) and overlapping $X=(3,1)_{4/3}$, $Y=(3,2)_{-5/6}$ (shaded in blue), assuming predominantly prompt $\eta_{\mathcal Q}$ production and no additional significant decay modes besides $\eta_{\mathcal Q} \to gg, \gamma\gamma$ (in full lines) or saturating their total decay width of $\sim 45$~GeV with hidden decay channels (in dot-dashed lines).  We observe that without additional decay modes or production mechanisms the preferred  region of $|R_1(0)|$ lies below the bottomonium regime for all four cases. As discussed above, color octet states are expected to be suppressed in this regime. Additonal $\eta_{\mathcal Q}^1$ decay modes, for example into lighter TC hadrons however can move the preferred region to higher values of $|R_1(0)|$. Finally we note that increasing $N_{\rm TC}$, considering additonal production mechanisms such as continuum production, or considering multiple states (see Sec.~\ref{sec:width}), pushes the preferred values of $|R_1(0)|$ further down.

\section{Excited States and the di-photon width}
\label{sec:width}

One interesting aspect of the ${\mathcal Q}$-onium system is the existence of excited states 
with equal quantum numbers that correspond to the radial excitations of the  system. We estimate the binding energy as,
\begin{equation}
\Delta M \sim  C_N {\alpha}_{\rm TC}  |R(0)|^{\frac 2 3}\,.
\end{equation} 
In the Coulomb regime (\ref{eq:hydrogen}) the splitting between ${}^1S_0$ and $2{}^1S_0$ for $N_{\rm TC}=3$ is given by,
\begin{equation}
\left(\frac {\Delta M}{M}\right)_{\rm Coul.}= \frac 1 6  {\alpha}_{\rm TC}^2   \,.
\end{equation} 
When non-perturbative effects become important a larger splitting (and correspondingly a larger signal) is obtained.
In Table~\ref{table:QCDeta} the mass splitting between (1S) and (2S) states in QCD are reported.
For the bottomonium in QCD for example the splitting is $6 \%$, much larger than indicated from the formula above ($\alpha_s(m_b)\approx 0.2$). 
This invites to interpret the apparent large width of the di-photon excess as preferred by ATLAS ($\Gamma/M=0.06$) as due to the presence of the di-photon excitation, $2^1 S_0$. 
The ratio of the prompt production cross-sections depends to leading order only on the values of the wave-functions at the origin.  In the Coulomb regime this is just 1/8 for $2^1 S_0/{}^1S_0$. 
Taking the charmonium and bottomonium cases in Table~\ref{table:QCDeta} as guidance, the $2^1 S_0$ rate could be only around a factor 2-3 smaller contributing in significant way to the total cross-section.  A bound state with binding properties similar to the bottomium $\eta_b$ would produce an apparent width of $6\%$. 
With the quantum numbers of $U$ the value of the wave-function in QCD would indicate a width into photons $\Gamma(\eta_{\mathcal Q}^1 \to \gamma\gamma)/M \sim 3 \times 10^{-6}$ that could be allowed with decays into lighter TC hadrons. 
Given the experimental resolution into photons it will be possible with more data to distinguish this from the scenario of a single broad resonance. 
Detecting several peaks with decreasing strength will be a clean signature of this scenario. 

In the regime where the splittings cannot be resolved the excited states contribute to the total cross-section enhancing it by an $\mathcal O(1)$ factor.
In addition we also predict the existence of higher spin particles. Spin-1 resonances analogous to $J/\Psi$ states of charmonium can be produced from $q\bar{q}$ initial states.
The $\psi^8$ gives a significant cross-section into di-jets, see Table~\ref{table:models}. 
The splitting among these states is expected to be small (1\% for bottomonium)  so that they will not appear as separate  resonances in the di-jet invariant mass distribution. 
The $\psi^1$ on the other hand decays into pairs of leptons with  a cross-section that could be measured with future data.
States with orbital angular momentum are more difficult to produce since their production cross-sections are proportional to derivatives of the wave-function at the origin and thus suppressed.

\section{Invisible decays}
So far we have assumed that  the lightest new fermions are the ones that make the $\mathcal Q$-onium.
The only states lighter than the di-photon resonance are then TC glueballs. If kinematically accessible the di-photon resonance could also decay 
into TC glueballs. These would decay back to SM through higher dimensional operators $G^2_{\rm TC} F_{SM}^2$ generated by loops of heavy fermions. 
The final state with 4 SM gauge bosons is a generic feature also of models with fermions in the confined regime that could be searched for at the LHC.

The scenario can be simply generalized by adding fermions with mass smaller than $M/2$, either lighter or heavier than $\Lambda_{\rm TC}$.  By far the safest possibility is that these are singlets.  
For $m< \Lambda_{\rm TC}$ and $n\ge 2$ the lightest particles are TC pions. The ones made of different species are stable and they constitute viable Dark Matter candidates 
whose relic abundance could also be thermally produced \cite{Redi:2016kip}.

The di-photon resonance can decay to such TC pions. The decay into TC hadrons can be estimated in perturbation theory as the decay into TC gluons that  will eventually hadronize,
\begin{equation}
\frac{\Gamma(\eta_{\mathcal Q}^1\to GG)}{\Gamma(\eta_{\mathcal Q}^1 \to g g)} \approx \frac 9 {32 I_3^2} \frac {N_{\rm TC}^2-1}{N_{\rm TC}^2}   \frac {\alpha_{\rm TC}^2}{\alpha_s^2}\,.
\end{equation}
One also finds,
\begin{equation}
\frac{\Gamma(\psi_{\mathcal Q}^1\to GGG)}{\Gamma(\psi_{\mathcal Q}^1 \to g g g)} \approx \frac {d_3^2}{N_{\rm TC}^2}  \frac{D_N}{D_3} \frac {\alpha_{\rm TC}^3}{\alpha_s^3}\,.
\end{equation}
For $\alpha_{\rm TC} > \alpha_s$ the decay into TC hadrons could be dominant. The final states are mostly TC pions.
Including this invisible decay from eq. (\ref{eq:signal}) the di-photon excess is reproduced for,
\begin{equation}
\frac {\Gamma(\eta_{\mathcal Q}^1 \to \gamma\gamma)}{M} \approx  10^{-6} \times \left(1+\frac{\Gamma_{GG}}{\Gamma_{gg}}\right)\,.
\end{equation}
Note that only $\eta_{\mathcal Q}^1$  can decay to TC hadrons. Such invisible decays of $\eta_{\mathcal Q}^1$    thus effectively imply a larger production cross-section of $\eta_{\mathcal Q}^8$.
Consequently it not possible to achieve a genuine large width in these models if color octet states are unsupressed, due to indirect constraints, particularly di-jets and photon-jet resonance searches. 

The lightest TC baryons are stable so they are also good dark matter candidates \cite{Antipin:2015xia}.  
Their cosmological stability is robustly guaranteed by the fact that the TC baryon number is broken by dimension 6 operators while stability of TC pions could be violated by dimension 5 operators. 
Interactions with the SM will be mediated by the $\mathcal Q$-onium. Such interactions are however strongly suppressed. For example in QCD $\eta_c$ decays into $p\bar{p}$ with a branching  $\sim 10^{-3}$.
TC baryons couple strongly to TC pions so they will be in thermal equilibrium with them. As a consequence if these are in thermal equilibrium with the SM  
the thermal relic abundance will be too small. Dark Matter as thermal relic could be reproduced in region of parameters where the annihilation cross-section 
of TC baryons is suppressed, for example when fermion masses are above $\Lambda_{\rm TC}$.

\section{Summary}

To conclude we can compare the $\mathcal{Q}$-onium system with other composite di-photon scenarios discussed in the literature.
When the fermions are lighter than $\Lambda_{\rm TC}$ the lightest states are TC pions.
For an irreducible SM representation the quantum numbers of the TC pions are identical to the ones of $\eta_{\mathcal Q}$ studied here and the singlet state (the $\eta'$) will couple to gluons and photons if the constituents carry color and electric charge, providing a perfect candidate for the di-photon excess with identical branching 
fractions into SM gauge bosons as $\eta_{\mathcal Q}^1$. 
Moreover the heavier spin-1 resonances $\rho$ will have the same quantum  numbers as the $\psi$ states. 
For the $\mathcal Q$-onium spin-0 and spin-1 particles are almost degenerate leading to stronger constraints from di-jets.

Even without new strong interactions a bound state would form just because of QCD interactions \cite{Kats:2009bv,Han:2016pab,Kats:2016kuz}. 
The main difference in this case is that the value of the wave-function that controls the decay rates is set by $\alpha_s$ without non-perturbative enhancements,
leading to smaller cross-sections.  Obtaining the required di-photon rate requires $Q=4/3$ or larger. Since the color octet state is not bound under QCD, this scenario can avoid the strongest constraints from dijet and jet-photon resonance searches discussed here. Note however that such QCD effects can be relevant in certain region of parameters also in our setup, enhancing in particular the singlet signal and weakening or eliminating the color octets.

The main prediction of the $\mathcal Q-$onium scenario is the presence of other resonances with the pattern sketched in Fig. \ref{fig:qonium}.
Color octet resonances with spin-0 and spin-1 can be copiously produced at the LHC and could be visible in the di-jet or jet-photon invariant mass distributions. 
Singlet spin-1 resonances could produce signals in di-leptons. Interestingly the large width suggested by ATLAS data can most easily be reproduced by the production of nearby radial excitations. 
Detecting such a pattern would be a clear smoking gun of $\mathcal Q$-onium. 
On the other hand, due to poorer di-jet invariant mass resolution, $\eta_{\mathcal Q}^1$, $\eta_{\mathcal Q}^8$ $\psi_{\mathcal Q}^8$ 
will likely not appear as individual resonances in di-jet searches.

\begin{acknowledgments}

JFK would like to thank the CERN TH Department for hospitality while this work was being completed and acknowledges the financial support from the Slovenian Research Agency (research core funding No. P1-0035).  MR is supported by the MIUR-FIRB grant RBFR12H1MW. We wish to thank Andrea Mitridate, Alessandro Strumia, Andrea Tesi and Elena Vigiani
for discussion on related subjects.

\end{acknowledgments}

\begin{appendix}

\end{appendix}


\begin{thebibliography}{99}        

\bibitem{exp}
ATLAS Collaboration, ``Search for new physics decaying to two photons'' ATLAS-CONF-2015-081;
CMS Collaboration, ``Search for new physics in high mass diphoton events in proton-proton collisions at 13 TeV,'' CMSPAS-EXO-15-004;
  CMS Collaboration,
  ``Search for new physics in high mass diphoton events in $3.3~\mathrm{fb}^{-1}$ of proton-proton collisions at $\sqrt{s}=13~\mathrm{TeV}$ and combined interpretation of searches at $8~\mathrm{TeV}$ and $13~\mathrm{TeV}$,''
  CMS-PAS-EXO-16-018.

\bibitem{Franceschini:2015kwy} 
  R.~Franceschini {\it et al.},
    arXiv:1512.04933 [hep-ph].
    
\bibitem{others}    
K. Harigaya, Y. Nomura, 
Phys. Lett. B754 (2016) 151 [arXiv:1512.04850]; 
Y. Nakai, R. Sato, K. Tobioka, 
[arXiv:1512.04924];
A. Pilaftsis, 
Phys. Rev. D93 (2016) 015017 [arXiv:1512.04931];
A. Belyaev, G. Cacciapaglia, H. Cai, T. Flacke, A. Parolini, H. Serodio, 
[arXiv:1512.07242]; 
L. Bian, N. Chen, D. Liu, J. Shu, 
[arXiv:1512.05759]. E. Molinaro, F. Sannino, N. Vignaroli, 
[arXiv:1512.05334] arXiv:1602.00475; Y.~Bai and J.~Osborne,  
JHEP {\bf 1511}, 036 (2015) [arXiv:1506.07110 [hep-ph]];
  P.~Draper and D.~McKeen,
  arXiv:1602.03604 [hep-ph].
    
\bibitem{Redi:2016kip} 
  M.~Redi, A.~Strumia, A.~Tesi and E.~Vigiani,
  arXiv:1602.07297 [hep-ph].
  
\bibitem{Craig:2015lra} 
  N.~Craig, P.~Draper, C.~Kilic and S.~Thomas,
  arXiv:1512.07733 [hep-ph]. 
  
\bibitem{Agrawal:2015dbf} 
  P.~Agrawal, J.~Fan, B.~Heidenreich, M.~Reece and M.~Strassler,
  arXiv:1512.05775 [hep-ph].  
  
\bibitem{Curtin:2015jcv} 
  D.~Curtin and C.~B.~Verhaaren,
  Phys.\ Rev.\ D {\bf 93}, no. 5, 055011 (2016)
  [arXiv:1512.05753 [hep-ph]].
  
\bibitem{alphascale}
The couplings controlling the bound state should be evaluated at the scale associated to the typical momentum of the $q\bar{q}$ pair. 
In the Coulomb regime this is the inverse of the Bohr radius $a_0^{-1}\sim \alpha_{\rm TC} M$.   
  
\bibitem{Novikov:1977dq} 
  V.~A.~Novikov, L.~B.~Okun, M.~A.~Shifman, A.~I.~Vainshtein, M.~B.~Voloshin and V.~I.~Zakharov,
  Phys.\ Rept.\  {\bf 41}, 1 (1978);

\bibitem{Appelquist:1978aq} 
  T.~Appelquist, R.~M.~Barnett and K.~D.~Lane,
  Ann.\ Rev.\ Nucl.\ Part.\ Sci.\  {\bf 28}, 387 (1978).

\bibitem{Brambilla:2004} 
  N.~Brambilla {\it et al.} [Quarkonium Working Group Collaboration],
  hep-ph/0412158.
  
\bibitem{Kim:2015zqa} 
  S.~Kim,
  Phys.\ Rev.\ D {\bf 92}, no. 9, 094505 (2015)
  [arXiv:1508.07080 [hep-lat]].

\bibitem{PDG}
Particle Data Group Collaboration,
{\em   ``Review of Particle Physics'',}
  Chin.\ Phys.\ C {38}, 090001 (2014).
  
\bibitem{Hagiwara:1990sq} 
  K.~Hagiwara, K.~Kato, A.~D.~Martin and C.~K.~Ng,
  Nucl.\ Phys.\ B {\bf 344}, 1 (1990).
  doi:10.1016/0550-3213(90)90683-5

\bibitem{Cirelli:2005uq} 
  M.~Cirelli, N.~Fornengo and A.~Strumia,
  Nucl.\ Phys.\ B {\bf 753}, 178 (2006)
  [hep-ph/0512090].
    
\bibitem{Kamenik:2016tuv} 
  J.~F.~Kamenik, B.~R.~Safdi, Y.~Soreq and J.~Zupan,
  arXiv:1603.06566 [hep-ph].

\bibitem{Aad:2015kna} 
  G.~Aad {\it et al.} [ATLAS Collaboration],
  Eur.\ Phys.\ J.\ C {\bf 76}, no. 1, 45 (2016)
  [arXiv:1507.05930 [hep-ex]].

\bibitem{Khachatryan:2015cwa} 
  V.~Khachatryan {\it et al.} [CMS Collaboration],
  JHEP {\bf 1510}, 144 (2015)
  [arXiv:1504.00936 [hep-ex]].
  
\bibitem{Zgamma} 
  The ATLAS collaboration,
  ATLAS-CONF-2016-010.

  \bibitem{Harlander:2005rq} 
  R.~Harlander and P.~Kant,
  JHEP {\bf 0512}, 015 (2005)
  [hep-ph/0509189].

\bibitem{Drees:1989du} 
  M.~Drees and K.~Hikasa,
  Phys.\ Rev.\ D {\bf 41}, 1547 (1990).

\bibitem{Fuks:2007gk} 
  B.~Fuks, M.~Klasen, F.~Ledroit, Q.~Li and J.~Morel,
  Nucl.\ Phys.\ B {\bf 797}, 322 (2008)
  [arXiv:0711.0749 [hep-ph]].

\bibitem{Chivukula:2011ng} 
  R.~S.~Chivukula, A.~Farzinnia, E.~H.~Simmons and R.~Foadi,
  Phys.\ Rev.\ D {\bf 85}, 054005 (2012)
  [arXiv:1111.7261 [hep-ph]].

\bibitem{Idilbi:2009cc} 
  A.~Idilbi, C.~Kim and T.~Mehen,
  Phys.\ Rev.\ D {\bf 79}, 114016 (2009)
  doi:10.1103/PhysRevD.79.114016
  [arXiv:0903.3668 [hep-ph]].
  
\bibitem{CMS:2015neg} 
  CMS Collaboration [CMS Collaboration],
  CMS-PAS-EXO-14-005;
  G.~Aad {\it et al.} [ATLAS Collaboration],
  Phys.\ Rev.\ D {\bf 91}, no. 5, 052007 (2015)
  [arXiv:1407.1376 [hep-ex]].
  
\bibitem{gammaj} 
ATLAS Collaboration, Phys. Lett. B 728 (2014) 562, arXiv:1309.3230;
ATLAS Collaboration, arXiv:1512.05910;  CMS Collaboration, Phys. Lett. B 738 (2014) 274, arXiv:1406.5171

\bibitem{Aad:2014cka} 
  G.~Aad {\it et al.} [ATLAS Collaboration],
  Phys.\ Rev.\ D {\bf 90}, no. 5, 052005 (2014)
  [arXiv:1405.4123 [hep-ex]].


 
 \bibitem{Aad:2015kna} 
  G.~Aad {\it et al.} [ATLAS Collaboration],
  Eur.\ Phys.\ J.\ C {\bf 76}, no. 1, 45 (2016)
  doi:10.1140/epjc/s10052-015-3820-z
  [arXiv:1507.05930 [hep-ex]].

  \bibitem{Aad:2014fha} 
  G.~Aad {\it et al.} [ATLAS Collaboration],
  Phys.\ Lett.\ B {\bf 738}, 428 (2014)
  doi:10.1016/j.physletb.2014.10.002
  [arXiv:1407.8150 [hep-ex]].
  
  \bibitem{Aad:2015fna} 
 S.~Chatrchyan {\it et al.} [CMS Collaboration],
  Phys.\ Rev.\ Lett.\  {\bf 111}, no. 21, 211804 (2013)
  Erratum: [Phys.\ Rev.\ Lett.\  {\bf 112}, no. 11, 119903 (2014)]
  doi:10.1103/PhysRevLett.111.211804, 10.1103/PhysRevLett.112.119903
  [arXiv:1309.2030 [hep-ex]];
  G.~Aad {\it et al.} [ATLAS Collaboration],
  JHEP {\bf 1508}, 148 (2015)
  [arXiv:1505.07018 [hep-ex]].

\bibitem{Czakon:2013goa} 
  M.~Czakon, P.~Fiedler and A.~Mitov,
  Phys.\ Rev.\ Lett.\  {\bf 110}, 252004 (2013)
  doi:10.1103/PhysRevLett.110.252004
  [arXiv:1303.6254 [hep-ph]];
  M.~Czakon and A.~Mitov,
  Comput.\ Phys.\ Commun.\  {\bf 185}, 2930 (2014)
  doi:10.1016/j.cpc.2014.06.021
  [arXiv:1112.5675 [hep-ph]].
  
\bibitem{Antipin:2015xia} 
  O.~Antipin, M.~Redi, A.~Strumia and E.~Vigiani,
  JHEP {\bf 1507}, 039 (2015)
  [arXiv:1503.08749 [hep-ph]].  
    
\bibitem{Kats:2009bv} 
  Y.~Kats and M.~D.~Schwartz,
  JHEP {\bf 1004}, 016 (2010)
  [arXiv:0912.0526 [hep-ph]].  
  
\bibitem{Han:2016pab} 
  C.~Han, K.~Ichikawa, S.~Matsumoto, M.~M.~Nojiri and M.~Takeuchi,
  arXiv:1602.08100 [hep-ph].

\bibitem{Kats:2016kuz} 
  Y.~Kats and M.~Strassler,
  arXiv:1602.08819 [hep-ph].
  
\end{thebibliography}
\end{document}